\documentclass[twocolumn]{aastex63}

\usepackage{amsmath}

\newcommand{\cii}{[C\,{\sc ii}]}
\newcommand{\civ}{C\,{\sc iv}}

\newcommand{\mgii}{Mg\,{\sc ii}}

\newcommand{\hi}{H\,{\sc i}}
\newcommand{\hii}{H\,{\sc ii}}
\newcommand{\hei}{He\,{\sc i}}
\newcommand{\heii}{He\,{\sc ii}}
\newcommand{\heiii}{He\,{\sc iii}}
\shorttitle{Testing the Lensing Hypothesis of Young Quasars}
\shortauthors{Yue et al.}
\graphicspath{{./}{figures/}}

\begin{document}

\title{Detecting and Characterizing Young Quasars. III. \\The Impact of Gravitational Lensing Magnification}

\correspondingauthor{Minghao Yue}
\email{myue@mit.edu}

\author[0000-0002-5367-8021]{Minghao Yue}
\affiliation{MIT Kavli Institute for Astrophysics and Space Research, 77 Massachusetts Ave., Cambridge, MA 02139, USA}

\author[0000-0003-2895-6218]{Anna--Christina~Eilers}\thanks{Pappalardo Fellow}
\affiliation{MIT Kavli Institute for Astrophysics and Space Research, 77 Massachusetts Ave., Cambridge, MA 02139, USA}

\author[0000-0003-3769-9559]{Robert A. Simcoe}
\affiliation{MIT Kavli Institute for Astrophysics and Space Research, 77 Massachusetts Ave., Cambridge, MA 02139, USA}

\author[0000-0002-5615-6018]{Sirio Belli}
\affiliation{Dipartimento di Fisica e Astronomia, Università di Bologna, Via Gobetti 93/2, 40129, Bologna, Italy}

\author[0000-0003-0821-3644]{Frederick B. Davies}
\affiliation{Max-Planck-Institut für Astronomie, Königstuhl 17, D-69117 Heidelberg, Germany}

\author{David DePalma}
\affiliation{MIT Kavli Institute for Astrophysics and Space Research, 77 Massachusetts Ave., Cambridge, MA 02139, USA}

\author[0000-0002-7054-4332]{Joseph F.\ Hennawi}
\affiliation{Department of Physics, University of California, Santa Barbara, CA 93106-9530, USA}
\affiliation{Leiden Observatory, Leiden University, Niels Bohrweg 2, NL-2333 CA Leiden, Netherlands}

\author[0000-0002-3407-1785]{Charlotte A. Mason}
\affiliation{Cosmic Dawn Center (DAWN)}
\affiliation{Niels Bohr Institute, University of Copenhagen, Jagtvej 128, 2200 København N, Denmark}

\author[0000-0002-8984-0465]{Julian B. Mu\~noz}
\affiliation{Department of Astronomy, The University of Texas at Austin, 2515 Speedway, Stop C1400, Austin, TX 78712, USA}
\affiliation{Center for Astrophysics $|$ Harvard {\&} Smithsonian, Cambridge, MA, USA}

\author[0000-0002-7524-374X]{Erica J. Nelson}
\affiliation{Department for Astrophysical and Planetary Science, University of Colorado, Boulder, CO 80309, USA}

\author[0000-0002-8224-4505]{Sandro Tacchella}
\affiliation{Department of Physics, Ulsan National Institute of Science and Technology (UNIST), Ulsan 44919, Republic of Korea}
\affiliation{Kavli Institute for Cosmology, University of Cambridge, Madingley Road, Cambridge, CB3 0HE, UK}
\affiliation{Cavendish Laboratory, University of Cambridge, 19 JJ Thomson Avenue, Cambridge, CB3 0HE, UK}





\begin{abstract}
We test the impact of gravitational lensing on the lifetime estimates of seven high-redshift quasars at redshift $z\gtrsim6$. 
The targeted quasars are identified by their small observed proximity zone sizes, 
which indicate extremely short quasar lifetimes $(t_Q\lesssim10^5 \text{ yrs})$. 
However, these estimates of quasar lifetimes rely on the assumption 
that the observed luminosities of the quasars are intrinsic 
and not magnified by gravitational lensing, 
which would bias the lifetime estimates towards younger ages.  
In order to test possible effects of gravitational lensing,
we obtain high-resolution images of the seven quasars
with the {\em Hubble Space Telescope (HST)}
and look for signs of strong lensing.
We do not find any evidence of strong lensing, 
i.e. all quasars are well-described by point sources, and no foreground lensing galaxy is detected. 
We estimate that the strong lensing probabilities for these quasars 
are extremely small $(\sim1.4\times10^{-5})$,
{{and show that weak lensing changes the estimated quasar lifetimes
by only $\lesssim0.2$ dex}.
We thus confirm that the short lifetimes of these quasars are intrinsic.}
The existence of young quasars indicates a high obscured fraction, radiatively inefficient accretion, and/or flickering light curves for high-redshift quasars. We further discuss the impact of lensing magnification on measurements of black hole masses and Eddington ratios of quasars. 
\end{abstract}



\section{Introduction} \label{sec:intro}

Quasars are the most powerful class of active galactic nuclei (AGNs)
and play substantial roles in the evolution of galaxies across cosmic time.
Luminous quasars have been discovered up to redshift $z\sim7.5$ 
\citep[e.g.,][]{mortlock11,wu15,banados18,wang19,yang20,wang21},
suggesting that supermassive black holes (SMBHs) with masses $M_\text{BH}\gtrsim10^9M_\odot$ 
already exist when the universe is less than 1 Gyrs old.
These quasars challenge our knowledge about 
the formation and growth of SMBHs in the early universe. 
It has been suggested that in order to produce a $10^9M_\odot$ SMBH at $z=7$,
we need either stellar-mass black hole seeds
($M_\text{BH}\sim10^2M_\odot$, e.g., the remnants of Population III stars) 
to accrete at super-Eddington rates, 
or massive black hole seeds with $M_\text{BH}\sim10^4M_\odot$
that are not trivial to explain \citep[e.g.,][]{inayoshi16, yang21}.
To date, the origin and growth history of these high-redshift quasars 
are still unclear.

High-redshift quasars are also powerful probes of 
the intergalactic medium (IGM) and the reionization process.
Quasars at $z\sim6$ have been used to constrain the opacity of the IGM 
via the Ly$\alpha$ forest \citep[e.g.,][]{fan06,eilers18,yang20igm}.
At $z\gtrsim7$, the damping wings of quasars enable measurements of the IGM
neutral fraction along individual line-of-sights \citep[e.g.,][]{banados18,yang20,wang20igm}.

Meanwhile, the opaque IGM at $z\gtrsim6$ offers unique opportunities 
of investigating the growth history of high-redshift SMBHs.
This task is done by measuring the sizes of quasar proximity zones.
Specifically, the ultraviolet (UV) photons from a quasar ionize the surrounding IGM,
generating a region around the quasar 
that has enhanced transparency to Ly$\alpha$ photons. 
This region, known as the proximity zone, can be probed
using the rest-frame UV spectrum of the quasar \citep[e.g.,][]{fan06,eilers17}.
The size of the proximity zone $(R_p)$ increases with
the luminosity and the age of the quasar (also referred to as the quasar's lifetime, $t_Q$).
It is thus possible to estimate the lifetime of high-redshift quasars
by measuring their proximity zone sizes.

In the past few years, we have measured the proximity zone sizes
of several tens of quasars and estimated their lifetimes
\citep[e.g.,][]{eilers17, eilers18young, eilers20, eilers21, davies20}.
These quasars show an average quasar lifetime of $t_Q\sim10^6$ yrs, consistent with other observations of high-redshift quasars \citep[e.g.,][]{chen18, bosman20, morey21, khrykin21}.
Interestingly, some quasars in the sample have extraordinarily small
proximity zones, indicating short quasar lifetimes of $t_Q\lesssim10^5$ yrs.
These very young quasars put unique constraints on 
the models of SMBH growth in the early universe.
\citet{eilers21} discussed several possible hypotheses to explain 
the extremely short lifetimes of these quasars,
including a long obscured phase of the quasar during which the SMBH can grow without ionizing the surrounding IGM \citep[see also][]{satyavolu22}, 
as well as extremely low radiative efficiency of the mass accretion that would allow the black hole to gain mass in short periods of time.

However, since the size of the quasars' proximity zones depend on their intrinsic luminosity, the observed short quasar lifetimes could also be explained if strong gravitational lensing magnifies their luminosity, implying that the quasars' intrinsic luminosity would be much lower. 
If the young quasars turn out to be strongly lensed, 
we may have overestimated the production rate of the ionizing photons
and thus underestimated the time needed for the quasars to ionize their proximity zones, i.e., the lifetimes of the quasars.

At this point, it is unclear if the young quasars found in our previous studies
are gravitationally lensed or not, which makes the interpretation of these quasars complicated. Although ground-based observations have found no signs of strong lensing for these quasars, it is still possible that these quasars are compact lensing systems that are unresolved in ground-based images.
One example of this is the quasar J0439+1634, a lensed quasar at $z=6.52$ with a small lensing separation of $\Delta \theta=0\farcs2$ and a large magnification of $\mu=51$ \citep[][]{fan19}. J0439+1634 is unresolved in ground-based imaging even with adaptive optics, and its lensing nature was confirmed only after the {\em Hubble Space Telescope (HST)} was able to resolve its lensing structure and detect the foreground deflector galaxy. J0439+1634 is also an excellent example of how lensing magnification affects quasar lifetime estimates. \citet{davies20} shows that
the estimated lifetime of J0439+1634 is $t_Q\approx10^3$ yrs before accounting for the magnification effect, which becomes $t_Q\gtrsim10^6$ yrs after correcting for the lensing magnification.


In this paper, we examine the lensing hypothesis for seven young high-redshift quasars using high-resolution {\em HST} images.
We look for signs of foreground deflector galaxies and multiple lensed images of the quasars, and estimate the probabilities for these quasars to be strongly lensed. In addition, we develop a method to quantify the impact of lensing magnification on quasar lifetime measurements, where we consider the impact of both weak and strong lensing. 
This paper is organized as follows. 
Section \ref{sec:sample} describes the sample of young quasars.
Section \ref{sec:hst} describes the high-resolution {\em HST} imaging and data reduction.
Section \ref{sec:lensing} presents the impact of lensing magnification on quasar lifetime measurements.
We discuss our results in Section \ref{sec:discussion} and conclude with Section \ref{sec:conclusion}. Throughout this paper, we use a flat $\Lambda$CDM cosmology with $\Omega_M=0.3$ and $H_0=70\text{ km s}^{-1}\text{Mpc}^{-1}$.

\section{The Young Quasar Sample} \label{sec:sample}

The sample of young quasars analyzed in this work is gathered from \citet{eilers18young}, \citet{davies20}, and \citet{eilers21}.
We refer the readers to \citet{eilers17,eilers18young,eilers20,eilers21} for details about quasar proximity zone size measurements and the quasar lifetime estimations. Here we summarize the basic ideas of these measurements.

\citet{eilers17} presents measurements of the proximity zone sizes of 34 quasars
using high S/N optical and near-infrared spectra. 
The proximity zone sizes are measured following the definition in \citet{fan06}.
Specifically, the quasar spectra are normalized by their intrinsic emission, 
which is estimated using principal component analysis (PCA) trained on low-redshift quasar spectra \citep[e.g.,][]{suzuki06,paris11,davies18_pca,bosman21}.
The normalized spectra are then smoothed with a 20{\AA} wide boxcar window
to measure the transmission level of the IGM.
The edge of the proximity zone is defined by the location where the
transmission level drops to 10\%.
\citet{eilers20} further improves the $R_p$ measurements of 12 quasars 
with small proximity zones using updated redshifts based on
{\cii} and {\mgii} emission lines.

To estimate the age of these quasars, 
\citet{eilers21} run radiative transfer (RT) simulations to model 
the dependence of $R_p$ on quasar lifetimes.
Specifically, \citet{eilers21} apply the 
one-dimensional RT simulation code from \citet{davies16}
on skewers from cosmological hydrodynamical Nyx simulation \citep[][]{almgren13,lukic15}.
The Nyx simulation was designed for
precision cosmological studies of the diffuse gas in the IGM, 
which includes $4096^3$ baryonic (Eulerian) grid
elements and dark matter particles.
As luminous quasars reside in massive dark matter haloes \citep[e.g.,][]{morselli14,onoue18,meyer22, kashino22},
\citet{eilers21} draw 1,000 skewers in random directions from the centers of
the most massive dark matter haloes in the Nyx simulation
to model the line-of-sight of observations.
The RT code assumes a ``lightbulb" model for the quasar
(i.e., the quasar turns on and maintains a constant luminosity), 
and computes the abundance of six particle species in the IGM, 
i.e., $e^-$, {\hi}, {\hii}, {\hei}, {\heii}, and {\heiii}.
Accordingly, \citet{eilers21} compute the transmission of the quasar spectrum 
and thus the proximity zone size of the quasar along each skewer.
The final product of the RT simulation is a distribution of
$R_p$ given the luminosity and the lifetime of a quasar.
By comparing the RT simulations to 
the observed proximity zone sizes of quasars,
\citet{eilers21} estimate the lifetime of 10 quasars at $5.7<z<6.5$.
Lifetimes of other quasars are estimated in \citet{eilers18young}, \citet{davies20} and \citet{andika20}
using the same method. 

\begin{figure}
    \centering
    \includegraphics[trim={0.5cm, 0.5cm, 0, 0}, clip, width=1\columnwidth]{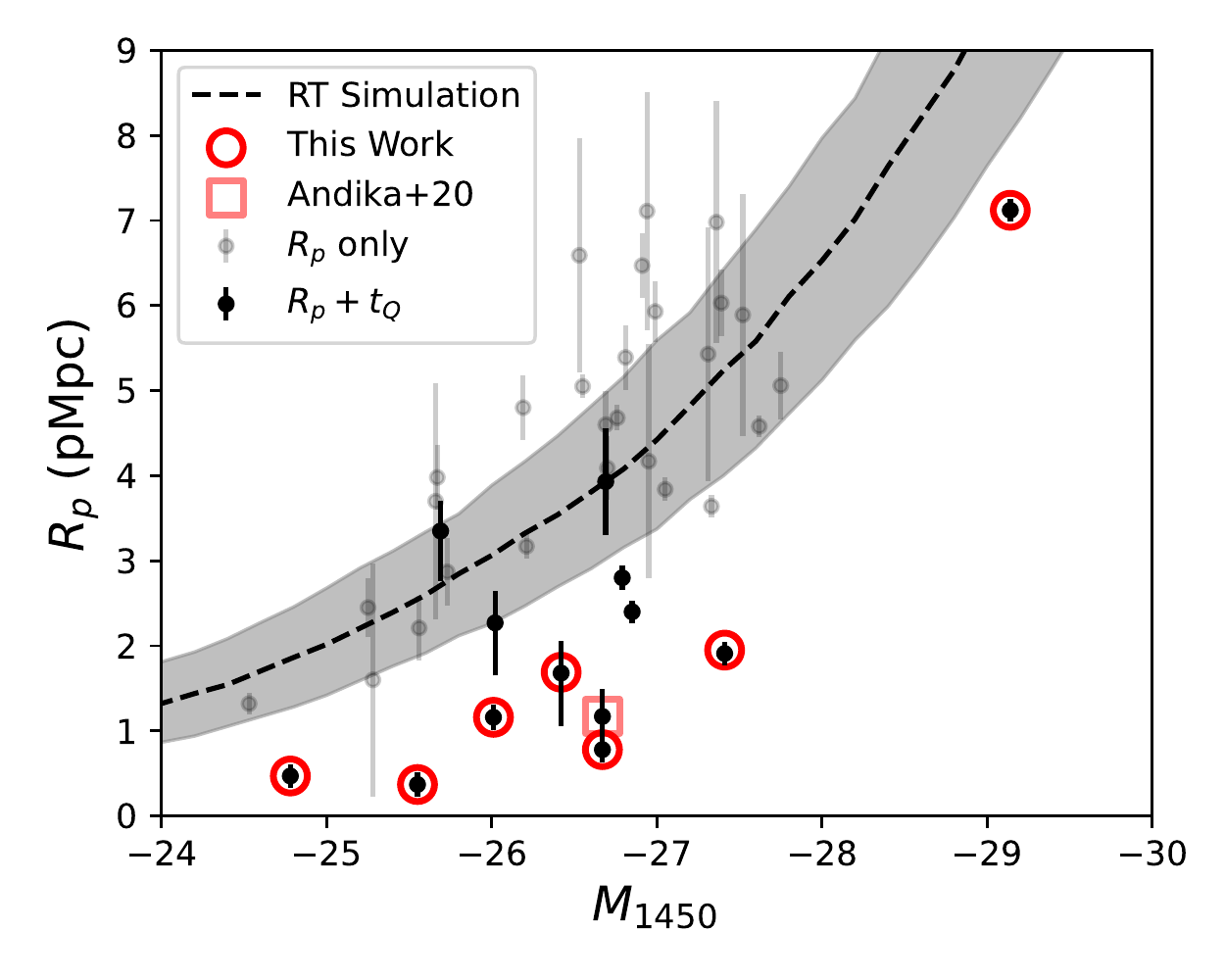}
    \caption{The $M_{1450}-R_p$ distribution of $z\sim6$ quasars from
    \citet{eilers17}, \citet{eilers20}, and \citet{andika20}. 
    Gray points are quasars with only $R_p$ measurements, 
    and black points are quasars with both $R_p$ and $t_Q$ estimates.
    The dashed line marks the median proximity zone size predicted by the RT simulation for $z\sim6$ quasars with
    $t_Q=10^6$ yrs, and the gray shaded area shows the $1\sigma$ error.
    The young quasars analyzed in this work are marked by red circles;
    these seven quasars have estimated $t_Q\lesssim10^{5}$ yrs,
    and have high-resolution images taken by {\em HST}.}
    \label{fig:sample}
\end{figure}

Figure \ref{fig:sample} presents the distribution of $M_{1450}$ 
(the absolute magnitude at rest-frame 1450{\AA})
and $R_p$ of quasars at $z\gtrsim6$ from the literature.
We also plot the $R_p$ distribution for quasars at $z\sim6$ from the RT simulation, which corresponds to a typical lifetime of $t_Q=10^6$ yrs \citep[e.g.,][]{khrykin21}.
Some quasars show proximity zone sizes smaller than the mean values of RT simulations by more than $1\sigma$, suggesting short lifetimes for these quasars.
However, the observed short lifetimes can also be explained by lensing magnification, as discussed in Section \ref{sec:intro}.
The aim of this work is to examine whether these quasars
are strongly lensed and investigate the impact of lensing magnification on quasar lifetime measurements.

\textcolor{black}{
The young quasar sample of this work consists of seven quasars with 
short estimated lifetimes ($t_Q\lesssim10^{5}$ yrs),
which are marked by red circles in Figure \ref{fig:sample}.
Two of these quasars have archival high-resolution {\em HST} images,
and we further obtain {\em HST} images for the remaining five quasars.
The information of these quasars are summarized in Table \ref{tbl:sample},
and the {\em HST} observations are described in Section \ref{sec:hst}
with more details.
We note that \citet{andika20} analyzed one quasar with a short estimated lifetime of
$\log t_Q\text{ (yrs)}=3.4$, which exhibits no signs of strong lensing 
and is not included in our sample.}


\begin{deluxetable*}{c|ccccccc}
\label{tbl:sample}
\tablecaption{The young quasar sample}
\tablewidth{0pt}
\tablehead{\colhead{Quasar} & \colhead{RA} & \colhead{Dec} &  \colhead{Redshift} & \colhead{$M_{1450}$\tablenotemark{1}} & \colhead{$R_p$\tablenotemark{2}} & \colhead{$\log t_Q$\tablenotemark{3}} & \colhead{Reference\tablenotemark{4}}\\
\colhead{} & \colhead{(hh:mm:ss.ss)} & \colhead{(dd:mm:ss.s)} & \colhead{} & \colhead{(mag)} &  \colhead{(proper Mpc)}  & \colhead{(yr)} & \colhead{}}
\startdata
\hline
PSO 004+17 & 00:17:34.47 & +17:05:10.7 & 5.8165 & -26.01 & $1.16\pm0.15$ & $3.6^{+0.5}_{-0.4}$ & \citet{eilers21}\\
SDSS J0100+2802 & 01:00:13.02 & +28:02:25.8 & 6.327 & -29.14 & $7.12\pm0.13$ & $5.1^{+1.3}_{-0.7}$ & \citet{davies20}\\
VDES J0330--4025 & 03:30:27.92 & --40:25:16.2 & 6.249 & -26.42 & $1.69^{+0.62}_{-0.35}$ & $4.1^{+1.8}_{-0.9}$& \citet{eilers21}\\
PSO J158--14 & 10:34:46.51 & --14:25:15.9 & 6.0681 & -27.41 & $1.95\pm0.14$ & $3.8^{+0.4}_{-0.3}$ & \citet{eilers21}\\
SDSS J1335+3533 & 13:35:50.81 & +35:33:15.8 & 5.9012 & -26.67 & $0.78\pm0.15$ & $3.0\pm0.4$ & \citet{eilers18young}\\
CFHQS J2100-1715 & 21:00:54.62 & --17:15:22.5 & 6.0806 & -25.55 & $0.37\pm0.15$ & $2.3\pm0.7$ & \citet{eilers21}\\
CFHQS J2229+1457 & 22:29:01.65 & +14:57:09.0 & 6.1517 & -24.78 & $0.47\pm0.15$ & $2.9^{+0.8}_{-0.9}$ & \citet{eilers21}\\\hline
\enddata
\tablenotetext{1}{The absolute magnitude at rest-frame $1450${\AA}.}
\tablenotetext{2}{The proximity zone size.}
\tablenotetext{3}{The quasar lifetime.}
\tablenotetext{4}{The reference from which the $R_p$ and $t_Q$ measurements are adopted.}
\tablecomments{All errors are $1\sigma$ errors.}
\end{deluxetable*}

\section{Testing the Strong Lensing Hypotheses with {\em HST} Imaging} \label{sec:hst}

We use high-resolution {\em HST} images to 
test the strong lensing hypothesis for the young quasars.
Each quasar is observed with a red filter and a blue filter.
The red filter covers long wavelengths 
where the quasar has prominent flux. 
The blue filter covers wavelengths shorter than the rest-frame Lyman limit,
where the quasar has no flux due to IGM absorption.
In the case of strong lensing, the red image will reveal  multiple lensed images of the quasar, 
and the blue image will detect the foreground lensing galaxy. 

Table \ref{tbl:image} summarizes the information of the observations.
Five of the seven quasars are observed by {\em HST} ACS/WFC
in the F555W and the F850LP filters (Proposal ID: 16756, PI: Eilers).
The other two quasars (SDSS J0100+2802 and VDES J0330-4025)
have archival observations by ACS/WFC in the F775W filter
and WFC3/IR in the F105W filter, respectively;
these two quasars do not have blue images below the quasars' Lyman limit taken by {\em HST}, and we use the $g-$band image from the DESI Legacy Imaging Survey
\citep[][]{legacy} as their blue images.
{J0100+2802 is also observed by ACS/WFC in the F606W filter,
which is also presented in this work.
The {\em HST} images are reduced using the \texttt{astrodrizzle} package
\citep{drizzle} following the standard procedure. }

\begin{figure*}
    \centering
    \includegraphics[width=0.49\textwidth, trim={1.2cm 1cm 1.2cm 0.4cm}, clip]{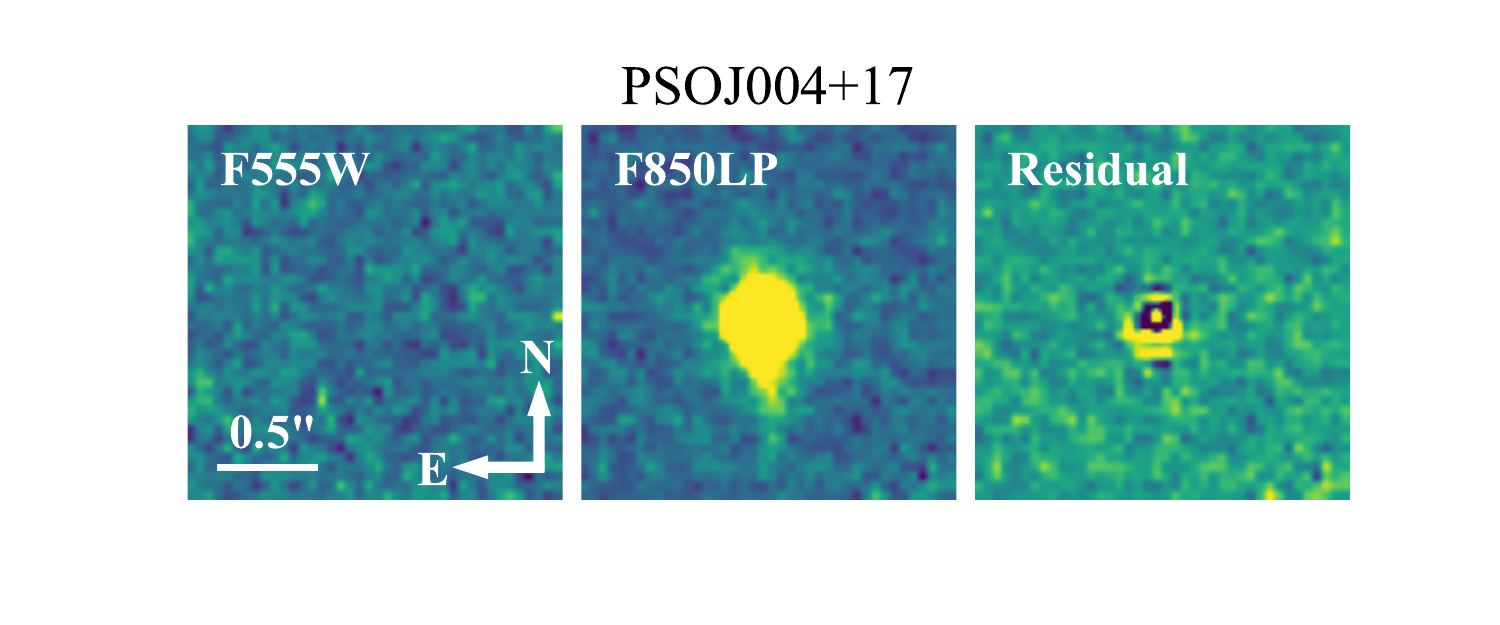}
    \includegraphics[width=0.49\textwidth, trim={1.2cm 1cm 1.2cm 0.4cm}, clip]{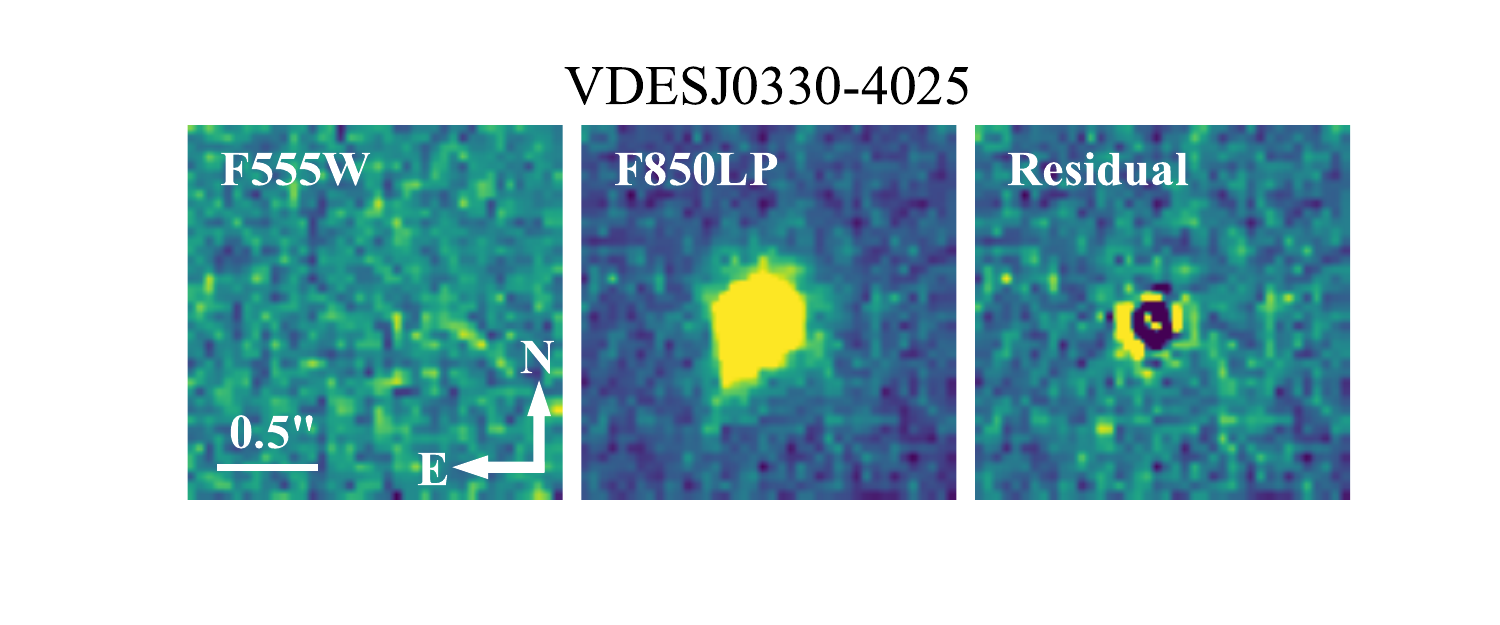}
    \includegraphics[width=0.49\textwidth, trim={1.2cm 1cm 1.2cm 0.4cm}, clip]{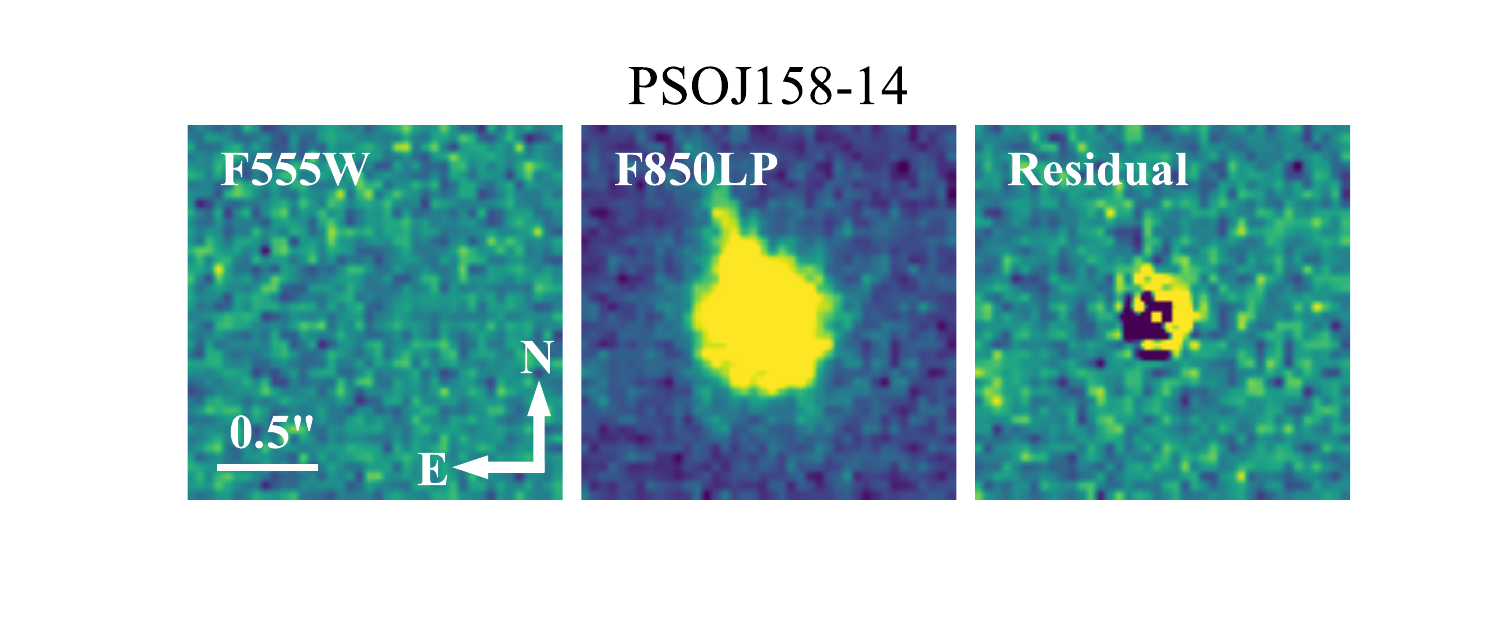}
    \includegraphics[width=0.49\textwidth, trim={1.2cm 1cm 1.2cm 0.4cm}, clip]{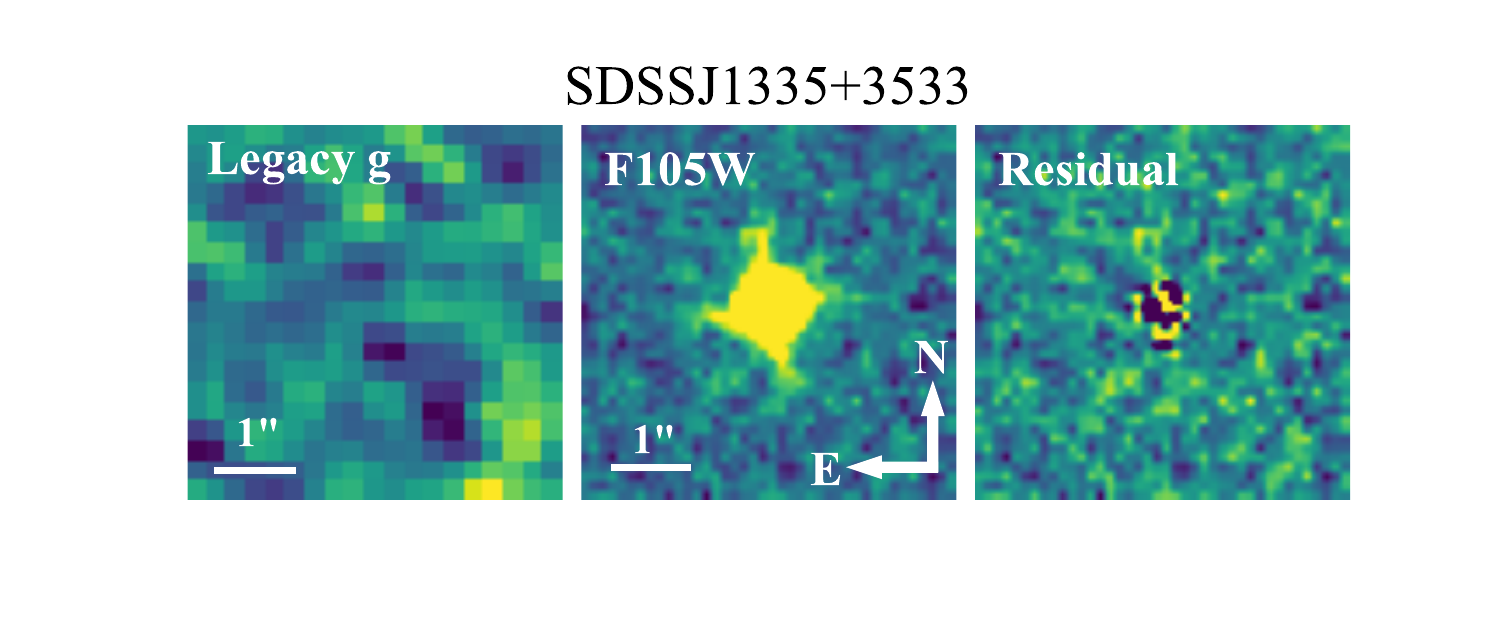}
    \includegraphics[width=0.49\textwidth, trim={1.2cm 1cm 1.2cm 0.4cm}, clip]{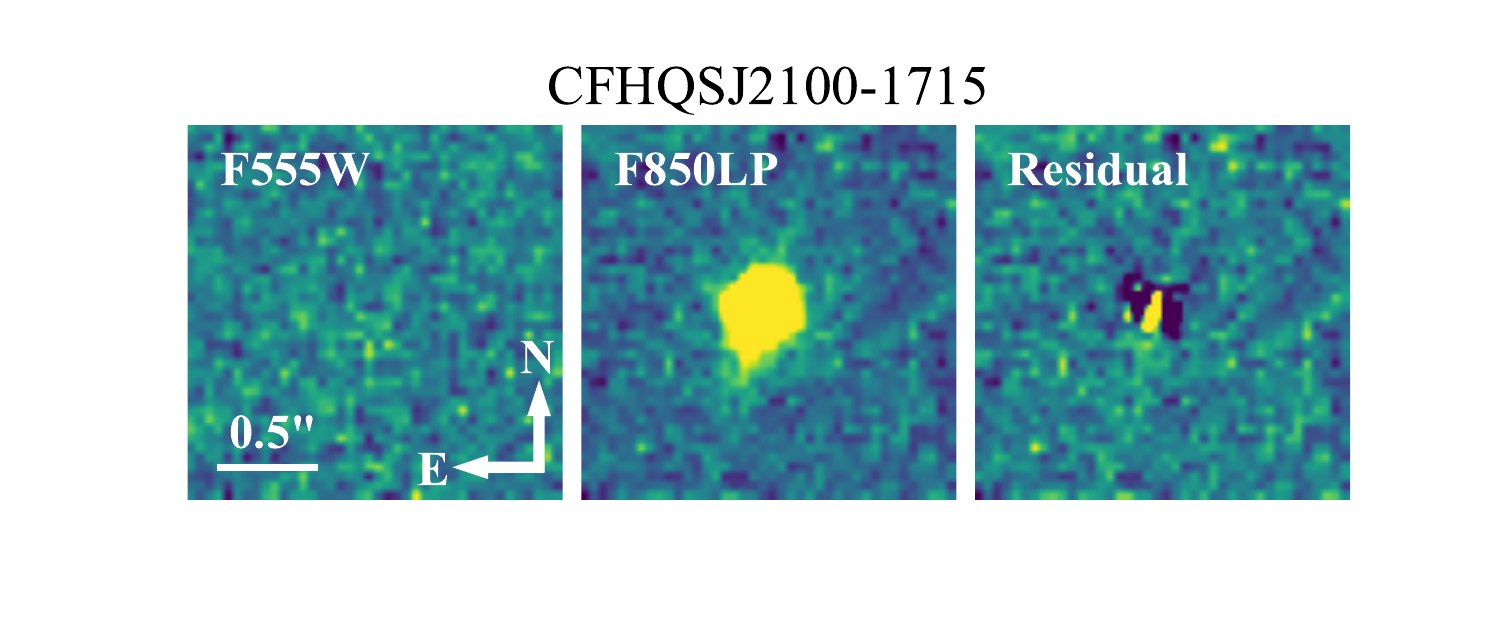}
    \includegraphics[width=0.49\textwidth, trim={1.2cm 1cm 1.2cm 0.4cm}, clip]{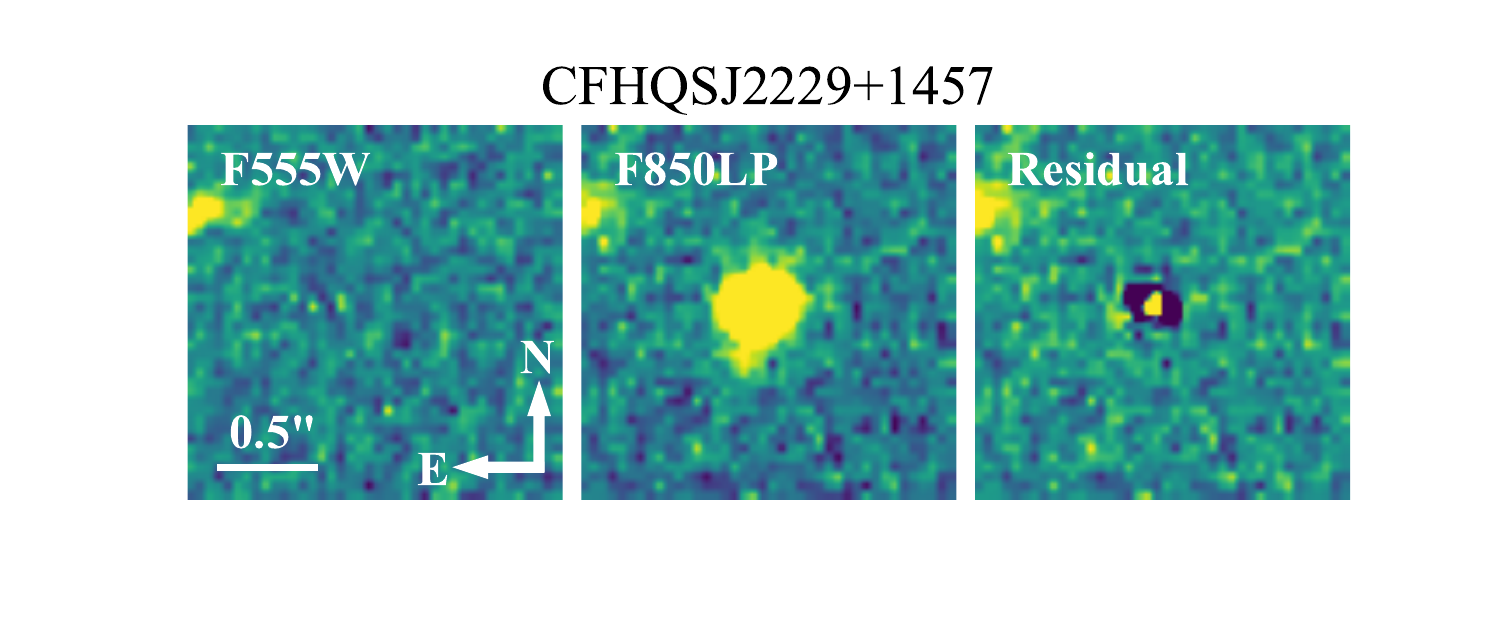}
    \includegraphics[width=0.7\textwidth, trim={1.2cm 1cm 1.2cm 0.4cm}, clip]{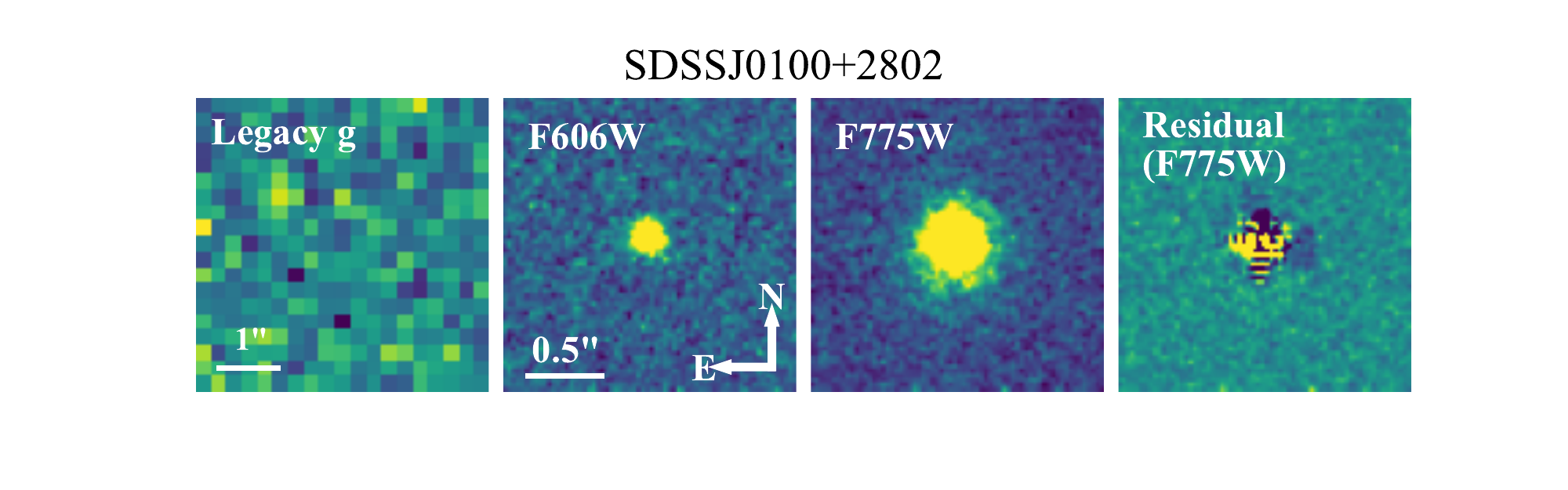}
\caption{The images of the young quasars. For each quasar, the panels from left to right show the image in a blue filter, the image in a red filter, and the  residuals of the red image after subtracting the PSF. All of the quasars are well-described by a single point source, and no foreground lensing galaxies is detected. There is no evidence of strong lensing for these quasars. Note that we include both F606W and F775W images for SDSS J0100+2802, which appears to be a single point source in both images.}
\label{fig:imaging}
\end{figure*}

Figure \ref{fig:imaging} presents the {\em HST} images,
where all the quasars appear to be point sources 
 and no foreground lensing galaxy is detected in these fields. 
In other words, there is no evidence of these quasars being strongly lensed. 


To further put quantitative constraints on possible lensing configurations,
for each quasar, we fit the red filter image as a point spread function (PSF).
We construct PSF models using IRAF task {\texttt{psf}} 
based on isolated stars in the field, 
and use {\em galfit} \citep[][]{galfit} to fit the quasar images as a single PSF.
Figure \ref{fig:imaging} shows the residual of the image fitting;
all the quasars are well-described by a single PSF
with no signs of a second lensed image.
We also try to fit the quasar images as two PSFs, 
which returns the same result as the single-PSF model
(i.e., the two PSF components are at the same position).

The {\em HST} images thus rule out the hypothesis that
the quasars are strongly lensed and have lensing separations larger than 
the resolution of the {\em HST} images.
Here we take the PSF full-width half maximum (FWHM; listed in Table \ref{tbl:image})
as the upper limit of the lensing separation,
{which is denoted by $\Delta\theta_\text{max}$ in the rest of this paper.} 
{The non-detection of the foreground lensing galaxy in the blue images
also indicate that these quasars are not strongly lensed;
however, as we will show in Section \ref{sec:analysis:lens},
the upper limits of the lensing separation give stronger constraints
on strong lensing probability compared to the blue images.}

In principle, it is still possible that these quasars are strongly lensed 
with small lensing separations that cannot be resolved by {\em HST}. 
Nevertheless, the {fraction} of strongly-lensed objects at $z\sim6$ 
{that have} lensing separations $\Delta \theta<0\farcs1$ is only $\sim1\%$, 
according to analytical models and mock catalogs of lensing systems \citep[e.g.,][]{yue22}.
As such, it is highly unlikely that the short observed lifetimes of these quasars 
are results of strong lensing.

\begin{deluxetable}{c|ccc}
\label{tbl:image}
\tablecaption{Imaging of the young quasars}
\tablewidth{0pt}
\tablehead{\colhead{Quasar} & \colhead{Filter} & \colhead{FWHM} & \colhead{Magnitude} \\
\colhead{} & \colhead{} & \colhead{$('')$} & \colhead{}}
\startdata
\hline
\multicolumn{4}{c}{Red Images (for background quasars)}\\\hline
PSO 004+17 & F850LP & $0\farcs10$ & 20.78\\
SDSS J0100+2802 & F775W & $0\farcs10$ & 21.30\\
VDES J0330--4025 & F850LP & $0\farcs13$ & 20.93\\
PSO J158--14 & F850LP & $0\farcs10$ & 19.69\\
SDSS J1335+3533 & F105W & $0\farcs10$ & 20.03\\
CFHQS J2100--1715 & F850LP & $0\farcs10$ & 21.67\\
CFHQS J2229+1457 & F850LP & $0\farcs10$ & 22.08\\\hline
\multicolumn{4}{c}{Blue Images (for foreground galaxies)}\\\hline
PSO 004+17 & F555W & $0\farcs1$& $>27.3$\\
SDSS J0100+2802 & Legacy $g$ & $1\farcs6$ & $>25.0$ \\
VDES J0330--4025 & F555W & $0\farcs1$ & $>27.4$\\
PSO J158--14 & F555W & $0\farcs1$ & $>27.3$\\
SDSS J1335+3533 & Legacy $g$ & $2\farcs1$ & $>24.2$\\
CFHQS J2100--1715 & F555W & $0\farcs1$ & $>27.2$\\
CFHQS J2229+1457 & F555W & $0\farcs1$ & $>27.3$\\\hline
\enddata
\tablecomments{The F555W, F775W, and the F850LP observations are taken with the 
{\em HST} ACS/WFC. The F105W image is taken with {\em HST} WFC3/IR.
The PSF FWHMs are estimated using stars in the field.
The magnitudes are all AB magnitudes, and the magnitude limits
 are $5-\sigma$ limits for point sources.}
\end{deluxetable}

\subsection{Additional Notes on SDSS J0100+2802}


SDSS J0100+2802 was initially reported by \citet{wu15}
as a {ultraluminous} quasar with a SMBH mass of $10^{10}M_\odot$. 
SDSS J0100+2802 was later observed by the
Atacama Large Millimeter/submillimeter Array
with a beam size of $0\farcs15$ \citep[][]{fujimoto20}.
The ALMA image of SDSS J0100+2802 exhibits four clumps, which are interpreted as 
the lensed images of the quasar host galaxy in \citet{fujimoto20}.
The fiducial lensing model suggested by \citet{fujimoto20}
has an image separation of $\Delta \theta=0\farcs2$ and a magnification of $\mu=450$.

SDSS J0100+2802 is also a target of the Guaranteed Time Observation program 
(Proposal ID: 1243, PI: Lilly)
of the {\em James Webb Space Telescope (JWST)}. 
\citet{eilers22} present the NIRCam F115W, F200W, and F356W imaging of SDSS J0100+2802,
finding no evidence of strong lensing with separation larger than $0\farcs05$.
The {\em JWST} observation is consistent with the {\em HST} images reported in this work, and rules out the lensing model suggested by \citet{fujimoto20}.
In the rest of this paper, we use $\Delta \theta_\text{max}=0\farcs05$
as the maximum possible strong lensing separation for SDSS J0100+2802.

\subsection{Additional Notes on CFHQS J2229+1457}

Figure \ref{fig:imaging} shows that there is a foreground object in the NE direction
that is $1\farcs03$ away from the quasar J2229+1457.
In the {\em HST} image, this object can be well described by 
 a S\'ersic profile with a half-light radius of $R_e=0\farcs4$,
 a S\'ersic index of $n=3.45$, and an axis ratio of $q=0.57$.
The magnitudes of this object is $m_\text{F555W}=25.2$ and $m_\text{F850LP}=23.8$.
Given its detection in the F555W image, this object must be a foreground object, which could introduce a magnification to the background quasar.

Without the spectra and the redshift of the foreground object,
we are not able to accurately calculate its contribution to 
the total magnification of the background quasar.
{However, we notice that the quasar is not multiply imaged,
i.e., the impact of the foreground galaxy can be described by 
weak lensing. In the Section \ref{sec:lensing}, we will analysis}
 the effect of both strong lensing and weak lensing
on quasar lifetime measurements. As such,
the potential impact of the foreground object near CFHQS J2229+1457 
is covered by the case of weak lensing. 


\section{The Effect of Lensing Magnification: A Probabilistic Analysis} \label{sec:lensing}

The observational constraints on lensing models
(i.e., the maximum possible lensing separation and the flux limit of the deflector galaxy)
are ususally used to derive the probability for the object to be strongly lensed
\citep[e.g.,][]{zlr22}.
However, quantifying the implication of the strong lensing probability
on quasar lifetime measurements is not straightforward.
In this work, we develop a probabilistic method to
quantify the impact of lensing magnification on quasar lifetime estimates.
We also consider the magnification from weak lensing
which is usually ignored in previous studies.
Note that the term ``strong lensing" means that the object is
multiply imaged in this work.

\subsection{Strong Lensing Probability} \label{sec:analysis:lens}

We start our analysis from
the {\em a priori} probability for an object to be strongly lensed,
also known as the strong lensing optical depth $\tau_m$.
The lensing optical depth describes the probability of a  source at a random position
to be strongly lensed by a foreground galaxy and
is determined by the population of deflector galaxies
\citep[e.g.,][]{wyithe02,wyithe11,yue22}:
\begin{equation}\label{eq:tm}
    \tau_m=\int_0^{z_s} dz_d \int_0^{+\infty} d\sigma \phi(\sigma, z_d) \frac{d^2V_c}{d\Omega dz_d} \pi \theta_E(\sigma, z_d, z_s)^2
\end{equation}
where $z_s$ and $z_d$ are the redshifts of the source and the deflector, 
$\frac{d^2V_c}{d\Omega dz_d}$ is the differential comoving volume,
$\phi(\sigma, z_d)$ is the deflector velocity dispersion function (VDF),
and $\theta_E$ is the Einstein radius of the deflector.

In this work, we use the parameterized VDF suggested by \citet{yue22}
that matches well with observed VDFs at $z\lesssim1.5$.
We use singular isothermal spheres (SISs) to describe the mass profile of deflectors,
which is widely used in modeling the population of lensing systems
\citep[e.g.,][]{wyithe11, mason15}.
The Einstein radius of an SIS deflector is given by 
$\theta_E = 4\pi\left(\frac{\sigma}{c}\right) ^2  \frac{D_{ds}}{D_{s}}$,
where $D_{s}$ and $D_{ds}$ are the angular diameter distances 
from the observer to the source and from the deflector to the source, 
and  $\sigma$ is the line-of-sight velocity dispersion.

It is useful to notice several properties of SIS lensing systems.
An SIS deflector generates two lensed images when
the angular separation between the background source and the deflector (denoted by $\beta$)
is smaller than the Einstein radius.
The two lensed images of the background source are separated by $\Delta\theta=2\theta_E$.
The lensing magnification is \textcolor{black}{determined by the 
 separation between the source and the deflector in the unit of $\theta_E$};
in other words, the magnification only depends on the lensing configuration
and does not rely on the  mass and the redshift of the deflector.

For the quasars in our sample,
the {\em HST} images rule out lensing models with large image separations.
After taking this constraint into consideration,
the strong lensing probabilities for these quasars are
\citep[see also][for a similar analysis]{zlr22}:

\begin{multline}\label{eq:plens}
    \tau_m(<\Delta \theta_\text{max})=\\
    \int_0^{z_s} dz_d \int_0^{\sigma_\text{max}} d\sigma \phi(\sigma, z_d) \frac{d^2V_c}{d\Omega dz_d} \pi \theta_E(\sigma, z_d, z_s)^2
\end{multline}
where $\sigma_\text{max}$ is the maximum velocity dispersion that can generate a compact lensing system allowed by the observation,
which is given by 
$\Delta\theta_\text{max}=2\theta_E(\sigma_\text{max}, z_d, z_s)$.

\textcolor{black}{Here, $\tau_m(<\Delta \theta_\text{max})$ gives 
the probability for a source at a random position
to be strongly lensed and has a lensing separation smaller than $\Delta \theta_\text{max}$.}
\footnote{\textcolor{black}{We notice that magnification bias can increase the {\em a posteriori} probability
of strong lensing \citep[e.g.,][]{wyithe11}. However, the magnification bias of $z\sim6$ quasars
is $\lesssim5$ for ordinary quasar luminosity functions and survey depths \citep{yue22},
which have no practical impact on our results as the values of $\tau_m(<\Delta \theta_\text{max})$
are exceedingly small.}}
Using Equation \ref{eq:plens}, 
we calculate the \textcolor{black}{value of $\tau_m(<\Delta \theta_\text{max})$}
for each quasar in our sample,
which are listed in Table \ref{tbl:lensprob}.
These values are extremely small $(\sim10^{-5})$,
indicating that the observed short lifetimes of the quasars 
are highly unlikely the results of strong lensing magnification.


In this work, we do not use the flux limit of the deflector galaxy to
constrain the strong lensing probability. 
Specifically, only faint and less massive galaxies are capable
of generating small-separation lenses that are unresolved by {\em HST}.
\textcolor{black}{
We estimate the flux of galaxies that have $\sigma<\sigma_\text{max}$ 
using the Faber-Jackson relation 
from previous observations \citep{bernardi03,focardi12}
and the galaxy spectra templates from \citet{brown14}.
We find that for galaxies at $z\gtrsim1$ 
\citep[the typical redshifts for deflector galaxies, e.g.,][]{collett13,mason15},
the F555W magnitudes are fainter than the 
image depths in Table \ref{tbl:image}.
In other words, the constraints we obtain from the image separation are more restrictive than the constraints based on the flux limit of a deflector galaxy in our observations.
We thus use the non-detection of the deflector galaxies as a cross-check
for our results that the young quasars do not exhibit signs of strong lensing.}
 
We also note that the estimated strong lensing probabilities
are subject to several systematic errors.
Specifically, the galaxy VDFs are not well-determined at $z\gtrsim1.5$,
and \citet{yue22} show that the uncertainties of galaxy VDFs
introduce a systematic error of $\sim30\%$ to the estimated lensing optical depth
for sources at $z\sim6$.
In addition, we use SIS models for deflectors
instead of more realistic elliptical mass distributions.
Nevertheless, the strong lensing probabilities are so small that 
the exact choices of deflector VDFs and lensing models
have essentially no impact on our analysis.
As we will show in Section \ref{sec:analysis:magnification},
weak lensing effects dominate the any magnification,
and the contribution of strong lensing is negligible.

\begin{deluxetable}{c|ccc}
\label{tbl:lensprob}
\tablecaption{Lensing Probabilities and the inferred quasar lifetimes}
\tablewidth{0pt}
\tablehead{\colhead{Quasar} & \colhead{$\Delta\theta_\text{max}$} & \colhead{$\tau_m(<\Delta\theta_\text{max})$}  & \colhead{$\log t_Q^\text{corr}$\tablenotemark{1}} \\
\colhead{} & \colhead{$('')$} & \colhead{} & \colhead{(yr)}}
\startdata
\hline
PSO004+17 & $0\farcs10$ & $1.43\times10^{-5}$ & $3.8^{+0.6}_{-0.3}$\\
J0100+2802 & $0\farcs05$ & $4.09\times10^{-6}$ & $4.7^{+2.0}_{-0.6}$\\
VDESJ0330-4025 & $0\farcs10$ & $1.48\times10^{-5}$ & $4.1^{+1.9}_{-0.5}$\\
PSOJ158-14 & $0\farcs10$ & $1.46\times10^{-5}$ &$3.7^{+0.4}_{-0.2}$\\
SDSSJ1335+3533 & $0\farcs13$ & $2.32\times10^{-5}$ & $3.1^{+0.4}_{-0.4}$\\
CFHQSJ2100-1715 & $0\farcs10$ & $1.46\times10^{-5}$ & $2.6^{+0.7}_{-0.8}$\\
CFHQSJ2229+1457 & $0\farcs10$ & $1.47\times10^{-5}$ & $3.2^{+0.8}_{-0.7}$\\\hline
\enddata
\tablenotetext{1}{The quasar lifetime with magnification distribution $P(\mu)$ 
taken into consideration.}
\end{deluxetable}

\subsection{Impact of Lensing Magnification on Quasar Lifetime Measurements} \label{sec:analysis:magnification}

{We have shown that the strong lensing probability
for the young quasars is only $\sim10^{-5}$.
However, the quantitative implication of lensing magnification
on the estimated quasar lifetimes is still unclear.
In particular, 
these quasars are subject to the magnification of weak lensing
even if they are not strongly lensed.
In this Section,
we derive the impact of lensing magnification (both strong and weak lensing)
on quasar lifetime measurements.}


Specifically, we compute $P(\mu)$
by marginalizing the cases of strong lensing and weak lensing,
following the method described in \citet{wyithe02}:
\begin{equation} \label{eq:pmu}
    P(\mu) = \tau_m(<\Delta\theta_\text{max})P_s(\mu)+
    [1-\tau_m(<\Delta\theta_\text{max})] P_w(\mu)
\end{equation}
where $P_s(\mu)$ and $P_w(\mu)$ are the distribution of magnification  
generated by strong lensing and weak lensing, respectively.
Recall that $\tau_m(<\Delta\theta_\text{max})$ is the probability
of strong lensing.

In this work, we adopt the weak lensing magnification distribution from \citet{mason15}.
\citet{mason15} use the code \texttt{Pangloss} \citep[][]{collett13} to compute the weak lensing magnification
of random line-of-sights in the Millennium simulation \citep[][]{springel05}.
We use the SIS lensing model to describe the magnification distribution for strong lensing,
i.e., $P_s(\mu)=8/\mu^3$ \citep[e.g.,][]{yue22}.
This distribution applies to $2<\mu<+\infty$, 
as the minimum strong-lensing magnification generated by an SIS lens is $\mu_\text{min}=2$.
Also note that this distribution is independent to the 
redshifts of the deflector and the background source, 
as well as the mass of the deflector. 

\begin{figure}
    \centering
    \includegraphics[width=1\columnwidth]{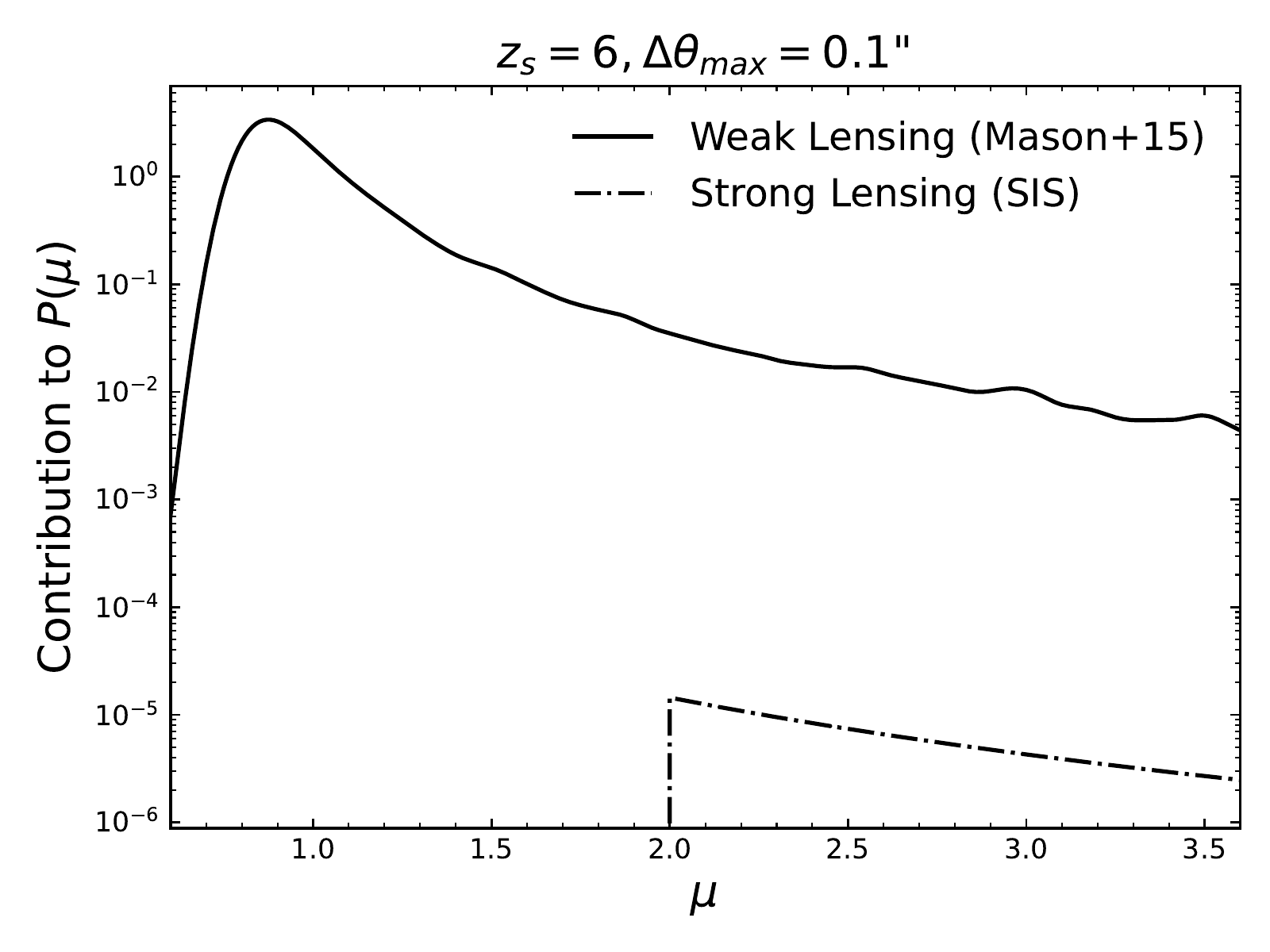}
    \caption{The contribution of weak lensing and strong lensing
    to the distribution of lensing magnification.
    This plot shows the case of a source at redshift $z=6$
    and a lensing separation limit of $\Delta\theta_\text{max}=0\farcs1$.
    The dotted dash and the solid lines illustrate the strong lensing term
    and the weak lensing term in Equation \ref{eq:pmu}.
    Given the small strong lensing probability $(\sim1.4\times10^{-5})$, 
    weak lensing dominates the marginalized distribution of lensing magnification.}
    \label{fig:pmu}
\end{figure}

Figure \ref{fig:pmu} shows the contribution of strong lensing and weak lensing
to $P(\mu)$
for a source redshift of $z=6$ and a lensing separation limit of $\Delta\theta_\text{max}=0\farcs1$.
The corresponding strong lensing probability is $\tau_m(<\Delta\theta_\text{max})=1.4\times10^{-5}$.
With such a small probability of strong lensing,
the contribution of weak lensing to $P(\mu)$ is about three orders of magnitude higher than that of strong lensing,
i.e., we have $P(\mu)\approx P_w(\mu)$. 
\textcolor{black}{Figure \ref{fig:pmu} demonstrates that the systematic uncertainties of 
$\tau_m(<\Delta\theta_\text{max})$ have little impact on the marginalized $P(\mu)$,
as we discussed in Section \ref{sec:analysis:lens}.}

We can now write down the marginalized distribution of $t_Q$,
\begin{equation} \label{eq:PtQ}
    P(t_Q)=\iint P(t_Q|M^\text{int}, R_p)P(M^\text{int})P(R_p) dM^\text{int} dR_p
\end{equation}
{where $M^\text{int}$ is the intrinsic (i.e., un-magnified) 
absolute magnitude of the quasar, and $P(M^\text{int})$ can be derived from $P(\mu)$
using the relation $M^\text{int}=M^\text{obs} + 2.5\log (\mu)$.
We follow the method in \citet{eilers22} to obtain $P(t_Q|M^\text{int}, R_p)$
and $P(R_p)$. Briefly speaking, $P(t_Q|M^\text{int}, R_p)$ is calculated using 
the RT simulations, and $P(R_p)$ is determined by the redshift uncertainties of the quasars.}

\begin{figure*}
    \centering
    \includegraphics[width=0.65\textwidth]{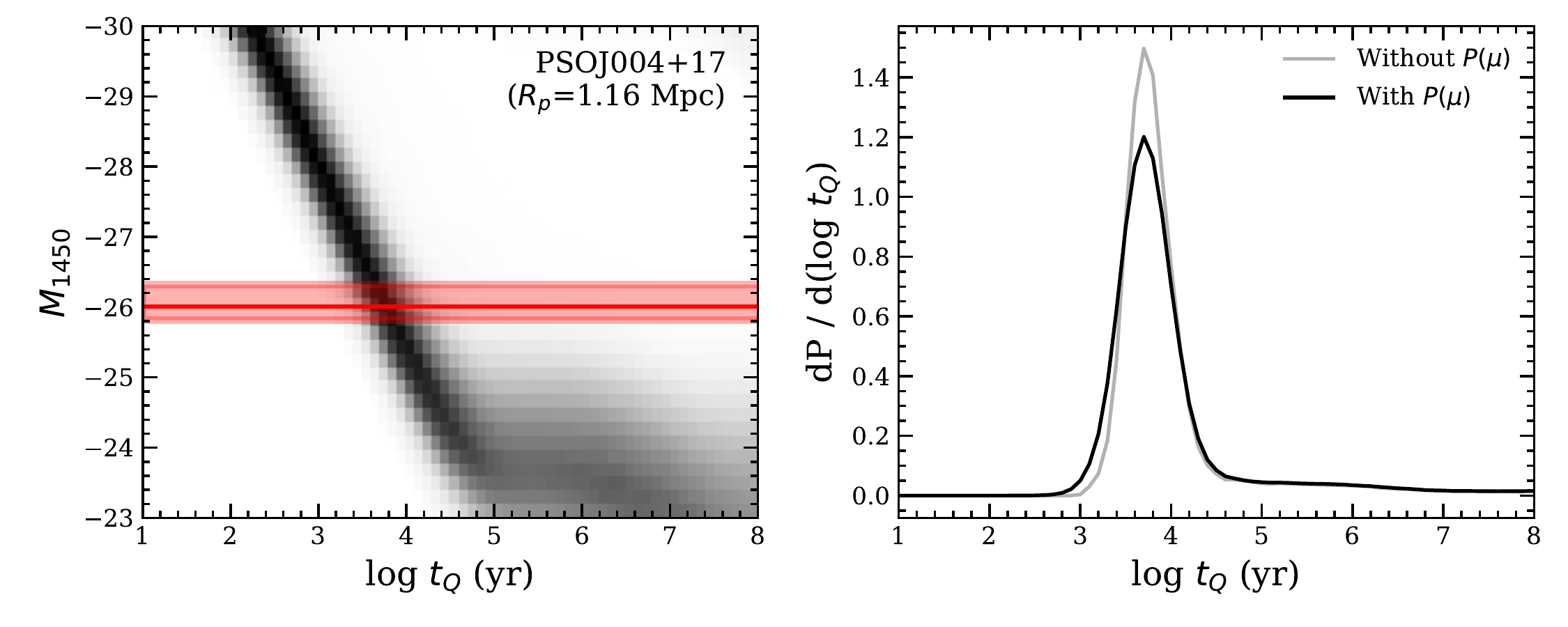}
    \includegraphics[width=0.65\textwidth]{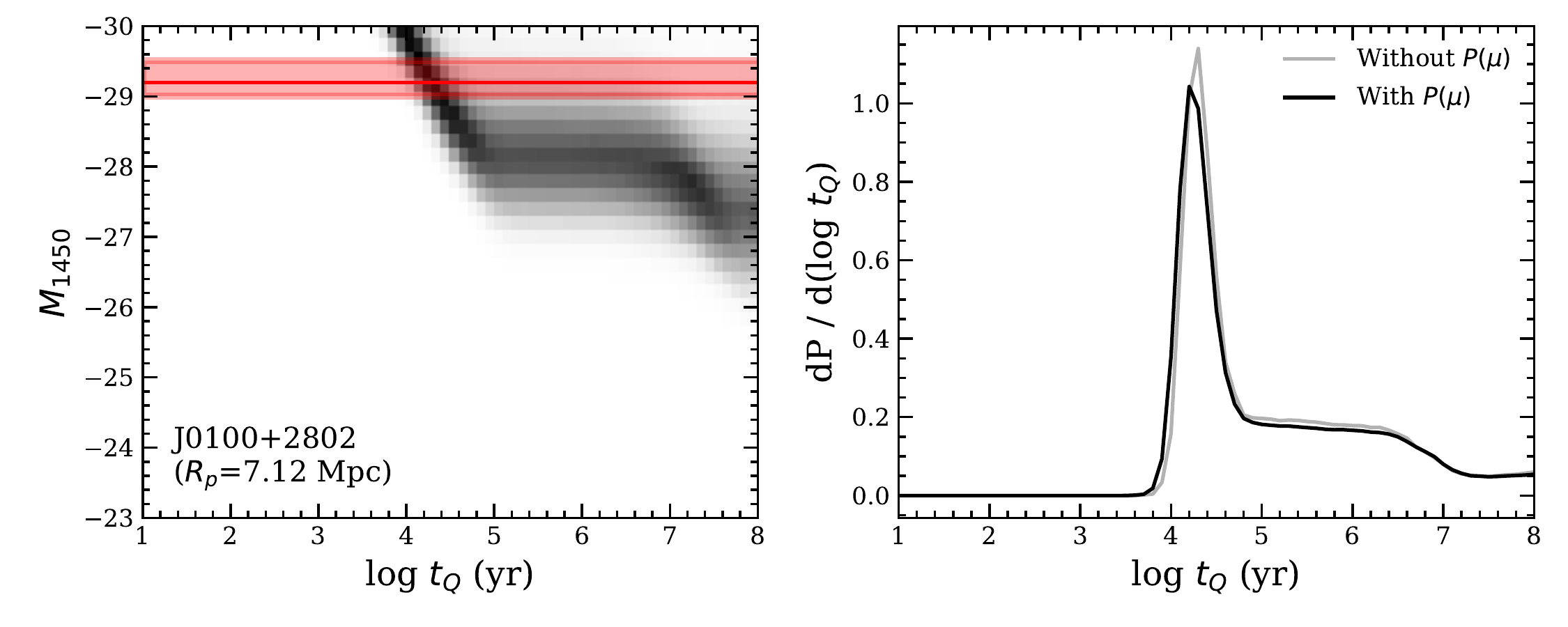}
    \includegraphics[width=0.65\textwidth]{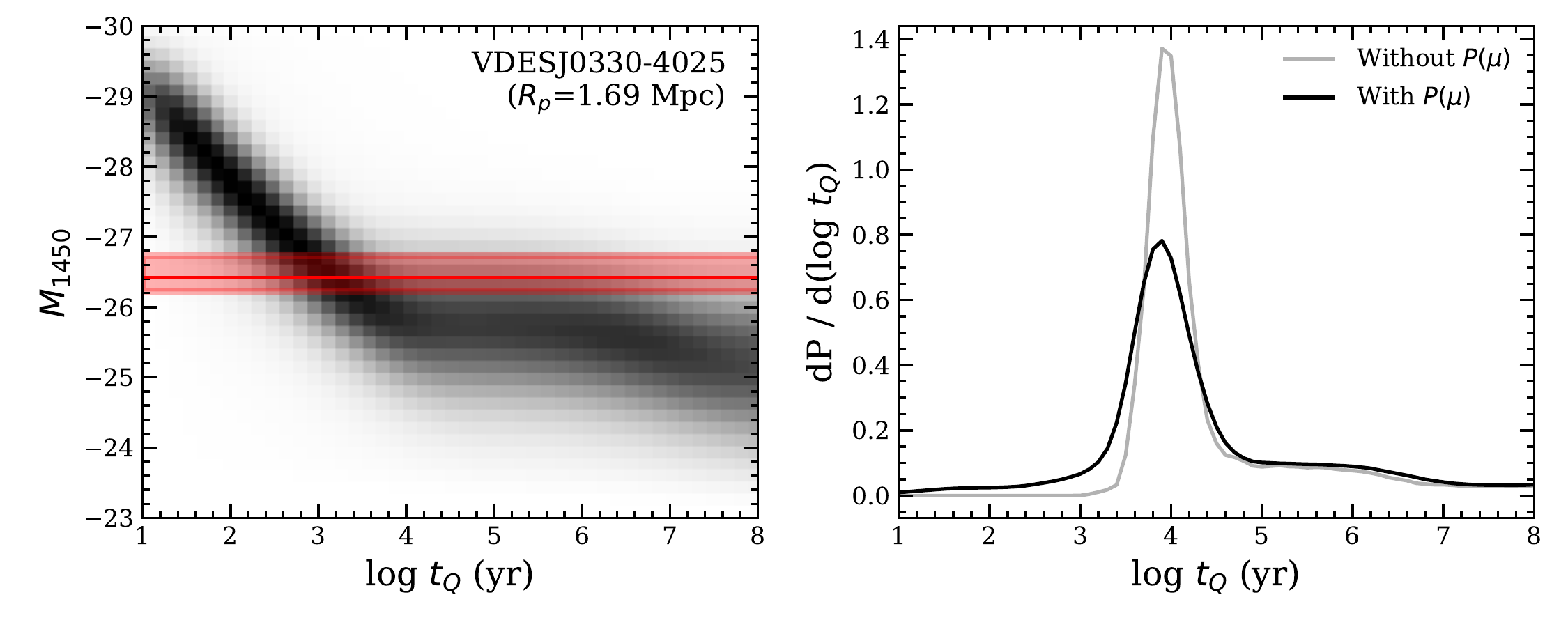}
    \caption{The impact of lensing magnification on quasar lifetime estimation.
    For each quasar, 
    the left panel presents the distribution of $M_{1450}$ and $t_Q$
    from the RT simulation, given the quasar's proximity zone size $R_p$.
    The red solid line marks the observed absolute magnitude of the quasar, 
    and the red shaded area shows the $1\sigma$ (68th percentile) range 
    of the intrinsic absolute magnitude after lensing magnification is considered. 
    The right panel shows the marginalized distribution of quasar lifetimes, $t_Q$,
    with and without considering the effect of lensing magnification.
    Despite some changes in the shape of $P(t_Q)$,
    lensing magnification has little impact on the estimated quasar lifetimes.}
    \label{fig:contour1}
\end{figure*}

Figure \ref{fig:contour1} illustrates the impact of lensing magnification
on quasar lifetime estimates.
The left column shows the distribution of $M_{1450}$ and $t_Q$
in the RT simulation, given the proximity zone sizes $R_p$ of each quasar.
The probability distribution of the intrinsic absolute magnitude
is marked by the red shaded area, which is determined by the observed absolute magnitude
(the red line) and $P(\mu)$. The right column shows the marginalized
distribution of $t_Q$ for each quasar calculated using Equation \ref{eq:PtQ}, 
with and without taking $P(\mu)$ into consideration. 
Despite some changes in the shape of the distribution, 
{$P(\mu)$ only shifts the mean values of estimated $t_Q$
by $\lesssim0.2$ dex.}
Figure \ref{fig:contour1} thus confirms that the quasars
in our sample are intrinsically young with $t_Q\lesssim10^5$ yrs, even
after taking into account the possible effects of lensing magnification.

\renewcommand{\thefigure}{\arabic{figure} (Continued)}
\addtocounter{figure}{-1}

\begin{figure*}
    \centering
    \includegraphics[width=0.65\textwidth]{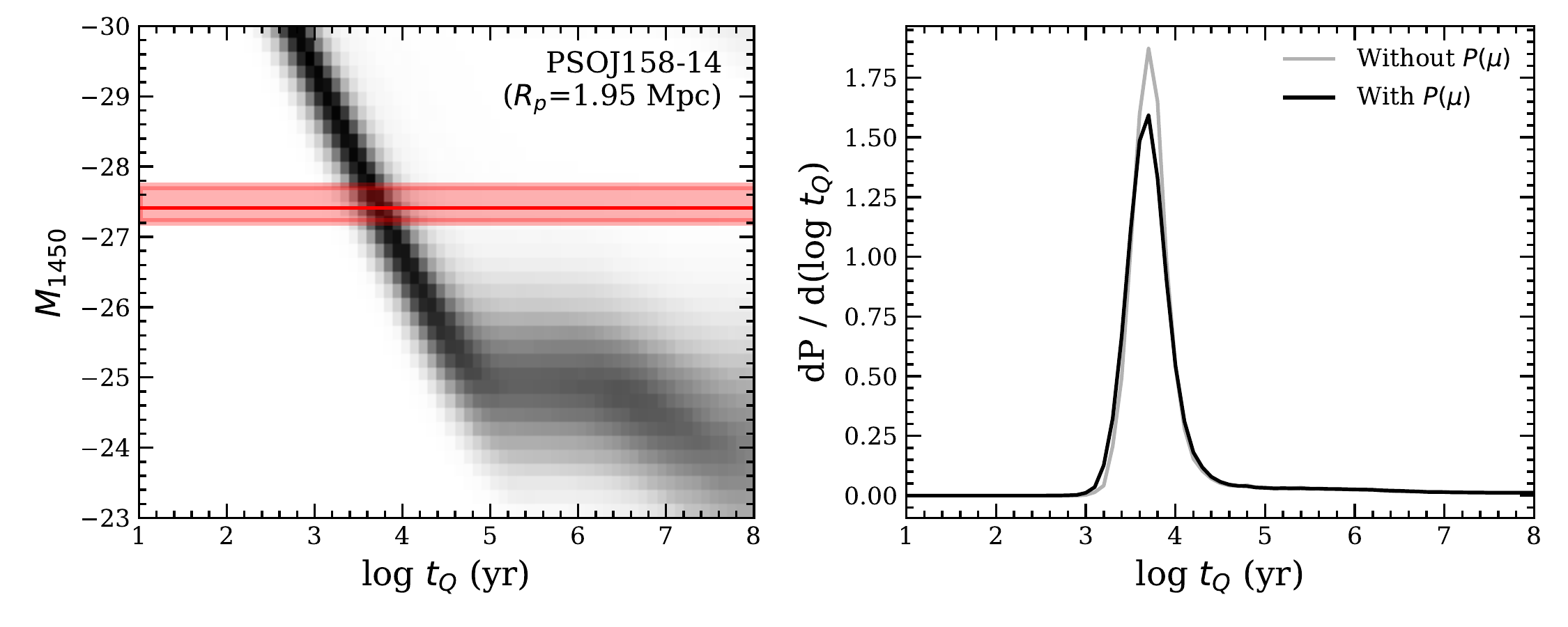}
    \includegraphics[width=0.65\textwidth]{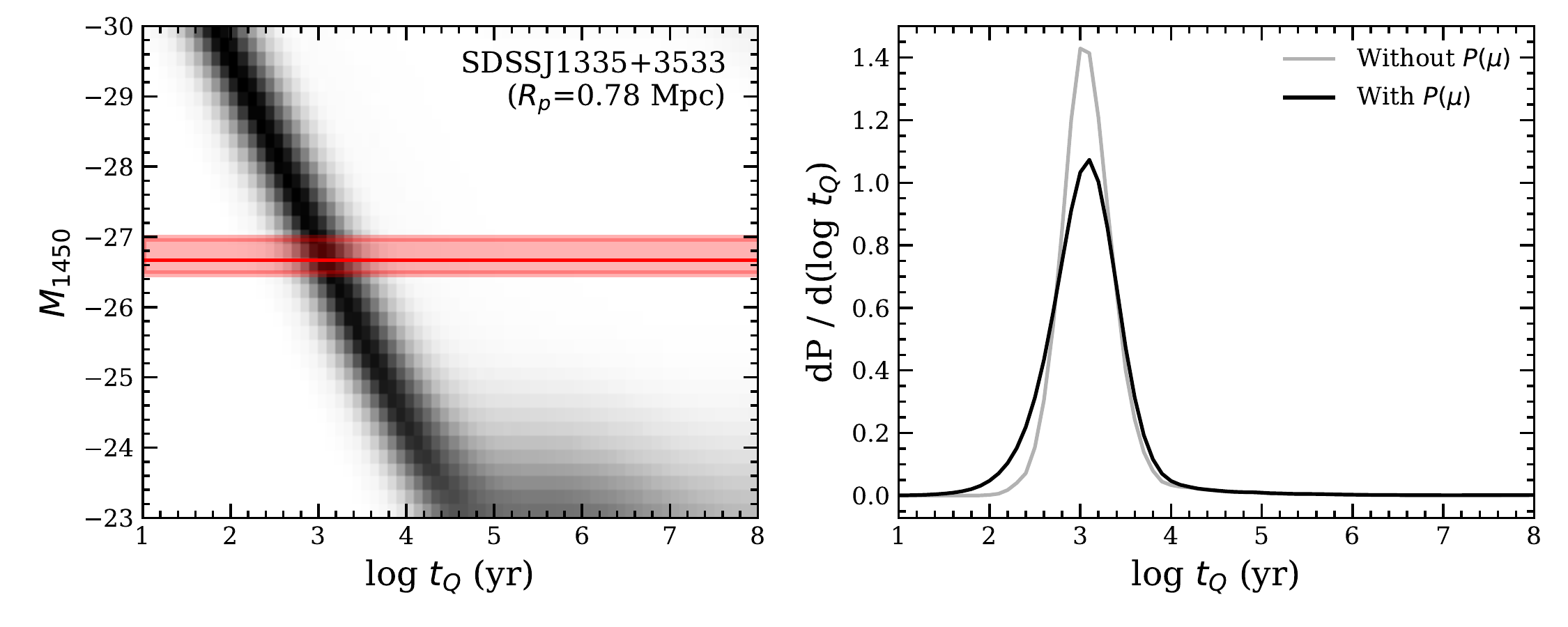}
    \includegraphics[width=0.65\textwidth]{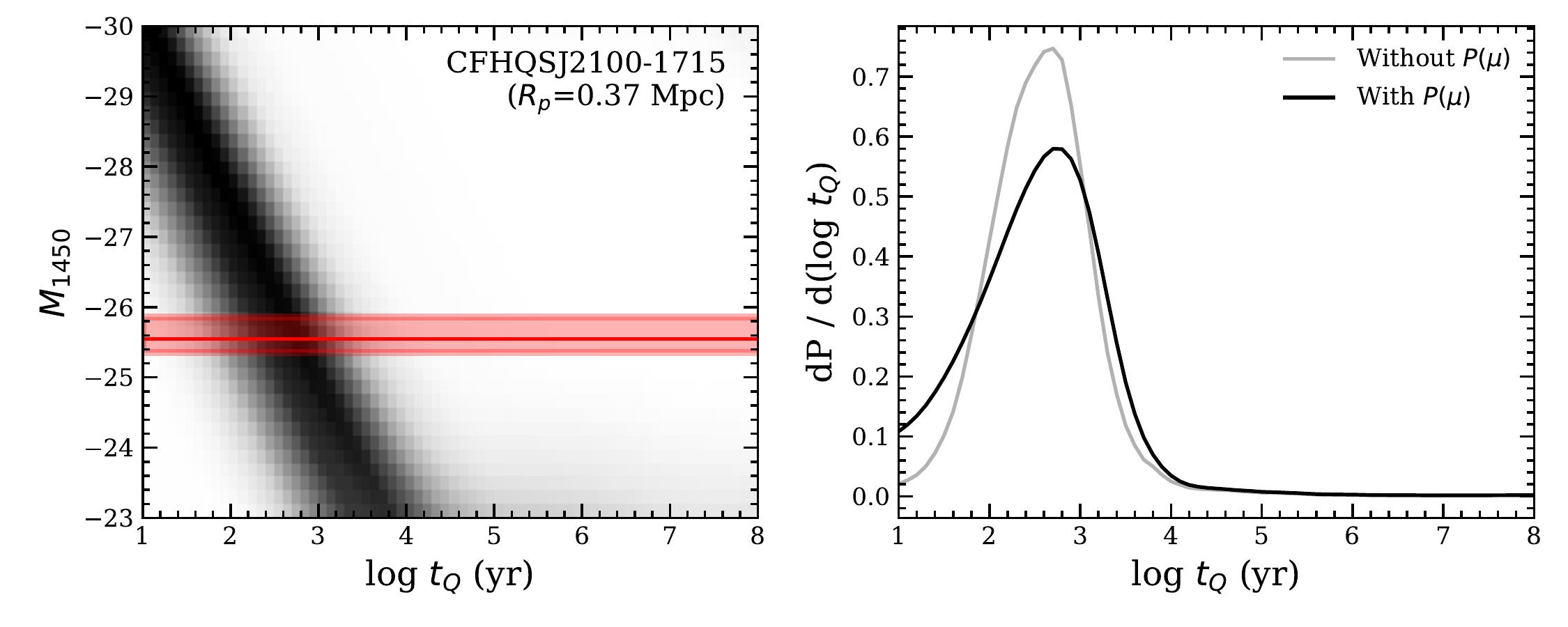}
    \includegraphics[width=0.65\textwidth]{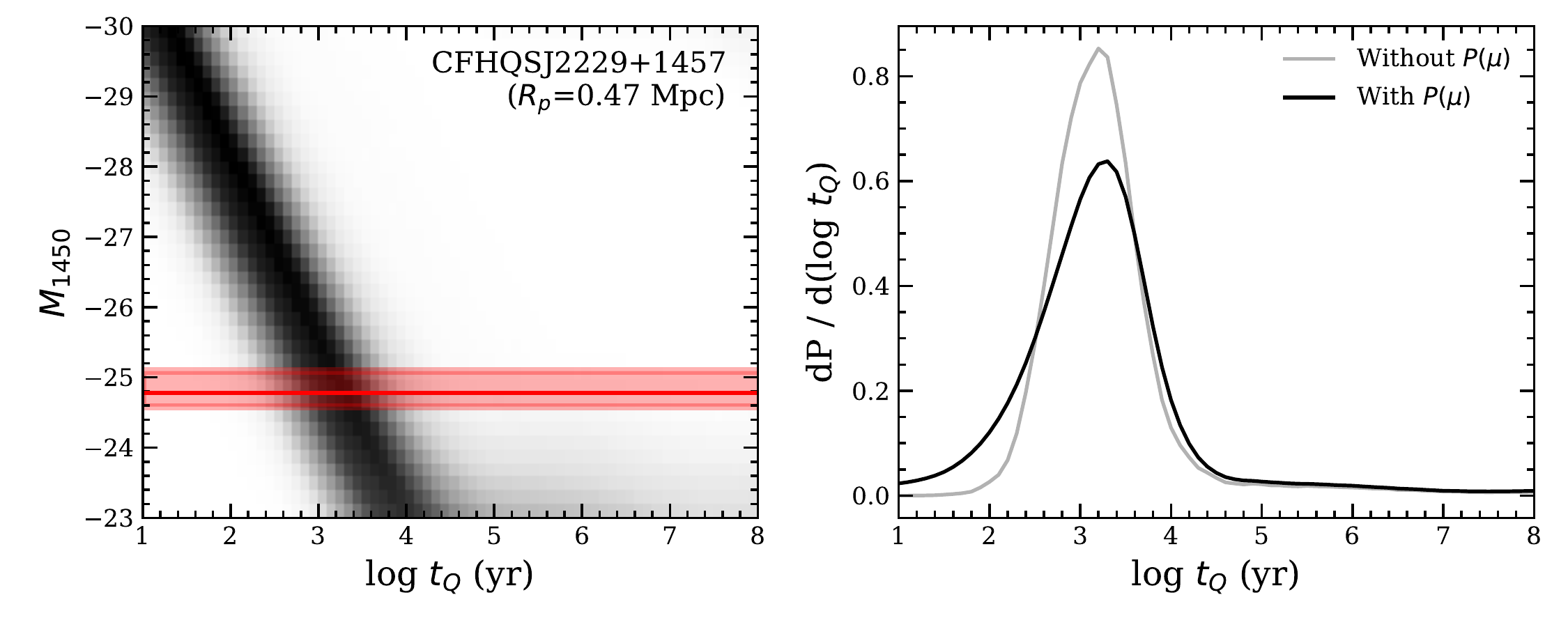}
    \caption{}
    \label{fig:contour2}
\end{figure*}

\renewcommand{\thefigure}{\arabic{figure}}

We end this Section by discussing the systematic errors of $t_Q$ estimates. 
\citet{eilers21} estimate the systematic errors
introduced by the RT simulation to be $\sigma_{\text{sys,}\log t_Q}\approx0.4$, 
which includes the diversity in quasar spectral energy distributions 
and reionization models.
{We suggest that the systematic uncertainty introduced by $P(\mu)$
is much smaller than the systematic uncertainty from the RT simulation. Specifically,}
the $P_w(\mu)$ given by \citet{mason15} is very close to other simulations
\citep[e.g.,][]{hilbert07}.
\citet{mason15} also calculate the $P_w(\mu)$ for a range of line-of-sight overdensities,
finding that the mean magnification differs by only $\lesssim 0.1$ dex.
Accordingly, we estimate the contribution of $P_w(\mu)$
to the systematic uncertainty of $t_Q$ to be $\lesssim 0.1$ dex.
As such, we still take 0.4 dex as the systematic errors of the $t_Q$ estimates.

\section{Discussion} \label{sec:discussion}


\subsection{The Implication of Young Quasars}

Section \ref{sec:lensing} suggests that 
the observed short lifetimes ($t_Q<10^{5}$ yrs) of the quasars are intrinsic
and are not results of lensing magnification.
\citet{eilers21} report that about $5\%$ of quasars at $z\gtrsim6$ 
have lifetimes $t_Q<10^{5}$ yrs; our results indicate that
the influence of lensing magnification on the observed young quasar fraction
is negligible.

Quasars with lifetimes $\lesssim10^5$ yrs 
put unique constraints on the AGN population and the
SMBH growth in the early universe. 
In the simple model where the black hole is accreting at a constant rate, 
the SMBH growth can be described by
\begin{equation}
    M_\text{BH}(t_Q)=M_\text{seed} \times \exp(t_Q/t_S)
\end{equation}
where $M_\text{seed}$ is the seed black hole mass, and $t_S$ is the Salpeter time 
\citep[][also known as the $e-$folding time]{salpeter64}  given by 
\begin{equation}
    t_S = 45\left(\frac{\epsilon}{1-\epsilon}\right)\left(\frac{L_\text{bol}}{L_\text{Edd}}\right)^{-1}\text{ Myr}
\end{equation}
where $\epsilon$ is the radiation efficiency of the accretion, which is about 0.1 for standard accretion disks \citep{ss73}. 
For seed black holes with $M_\text{seed}\sim10^2M_\odot$
(e.g., the remnants of Pop III stars),
we need $\epsilon\sim10^{-4}$ to form a SMBH with $M_\text{BH}=10^9M_\odot$
within $10^5$ yrs assuming $L_\text{bol}\approx L_\text{Edd}$.
{Such a low accretion efficiency is hard to achieve
even for hyper-Eddington accretion disks with Eddington ratios $\lambda_\text{Edd}>5\times10^3$}
\citep[e.g.,][]{inayoshi16}.

There are two viable explanations for the quasars with extremely short lifetimes.
First, the quasars might have experienced an obscured phase
where the SMBH is actively accreting material but not ionizing the surrounding IGM.
This picture is consistent with hydrodynamical simulations 
\citep[e.g.,][]{dm05, hopkins08}, which suggest that merger-triggered AGN
evolve from a UV obscured to an unobscured phase.
The fraction of accretion time in the UV obscured phase during the entire accretion history of SMBHs can be estimated by the obscured fraction of AGNs.
A large obscured fraction of high-redshift AGNs alleviates 
the difficulty of forming a quasar with a short unobscured, UV luminous lifetime \citep{davies19, satyavolu22}.
Observations have suggested a high obscured fraction 
of $\sim80\%$ \citep[e.g.,][]{vito18} for luminous AGNs.
\citet{endsley22} recently reported a heavily obscured hyperluminous AGN at $z=6.83$ 
in the 1.5 deg$^2$ COSMOS field, suggesting that the obscured fraction of high-redshift luminous quasars 
might be as high as $\sim99.5\%$. 
In the future, a complete sample of AGN is needed to accurately and correctly measure the AGN obscured fraction, which requires multi-wavelength surveys from X-ray to radio \citep[e.g.,][]{lyu22}.

Second, the quasar lifetime estimates in this work assume a light-bulb lightcurve
for the quasars, i.e., the quasar activity only turns on once and never turns off.
In contrast, SMBHs may have multiple periods of quasar activity,
which offers another explanation for the small proximity zones.
Specifically, if the time separation between two periods of quasar activity is sufficiently large, the IGM will become opaque to Ly$\alpha$ photons  before the second activity starts due to recombination, even if the first activity episode had ionized the surrounding IGM.
In other words, a SMBH can gain mass via previous phases of active accretion,
while only the most recent quasar activity is responsible for
the formation of the proximity zone of quasars at $z\sim6$.
This effect is analyzed in detail by \citet{davies20flickering} and \citet{satyavolu22}, 
who show that the small proximity zone sizes of quasars can be produced by 
a ``flickering" lightcurve (i.e., the quasar regularly turns on and off periodically).
This picture agrees with recent phenomenological models of high-redshift SMBH populations \citep[e.g.,][]{li22kiaa}, which suggest that  quasars at $z\sim6$ have experienced multiple periods of active accretion.
It is worth noticing that, even with flickering lightcurves,
the existence of high-redshift SMBHs at $z\gtrsim6$ still favors
a high obscured AGN fraction of $\gtrsim70\%$, as argued by \citet{satyavolu22}.

Based on the above considerations, we argue that the existence of quasars
with estimated lifetimes $t_Q\lesssim10^5$ yrs is consistent with the picture
where these quasars have experienced UV-obscured black hole growth, 
and might have had several periods of active accretion
prior to the current quasar activity,
possibly with radiatively inefficient ``super-Eddington'' accretion disks. 

\subsection{The Impact of Lensing Magnification on Quasar Property Measurements}

In addition to the quasar lifetimes and the trivial case of the quasars' luminosities,
lensing magnification also affects the measurements of other quasar properties.
Here we discuss two important examples of such properties,
i.e., the SMBH mass and the Eddington ratio of quasars.


The SMBH masses of quasars are often measured using the so-called
``single-epoch virial estimators" \citep[e.g.,][]{vp06,vo09}, 
which assume that the widths of the broad emission lines
originate from the virialized motion of the quasar's broad line region.
Specifically, the black hole mass is calculated using the FWHM of broad emission lines
(e.g., H$\alpha$, H$\beta$, \mgii, \civ) and the continuum luminosity:
\begin{equation} \label{eq:mbh}
    \log M_\text{BH} = a\log\text{FWHM}+b\log \lambda L_\lambda + c
\end{equation}
with the fiducial parameter values being $a=2$ and $b=0.5$.
Note that the FWHM of emission lines is not affected by lensing magnification.
Consequently, the apparent (i.e., without correcting for lensing magnification)
SMBH mass scales as $M_\text{BH}\propto\mu^{0.5}$ \citep[see also][]{fan19}. 

The Eddington ratio of a quasar is defined as
the ratio between its bolometric luminosity and the Eddington luminosity, i.e., 
\begin{equation} \label{eq:Redd}
    \lambda_\text{Edd}=\frac{L_\text{bol}}{L_\text{Edd}}=\frac{L_\text{bol}}{1.26\times10^{38} \text{ erg s}^{-1} \times (M_\text{BH}/M_\odot)}
\end{equation}
Since the apparent SMBH mass $M_\text{BH}$ is proportional to $\mu^{0.5}$,
according to Equation \ref{eq:Redd}, 
the apparent Eddington ratio of quasars also scales as $\lambda_\text{Edd}\propto\mu^{0.5}$.


Understanding the impact of lensing magnification on these quasar properties
is important in the studies of the SMBH population and evolution.
In particular, the luminosity functions, the SMBH mass functions and
the Eddington ratio distributions play critical roles
in the phenomenological models of SMBHs \citep[e.g.,][]{wu22jiang,li22kiaa}.
The impact of lensing magnification (especially from weak lensing)
should be correctly taken into account in such studies.

Meanwhile, SMBH masses and Eddington ratios provide useful tools
in surveys of strongly lensed quasars.
In particular, lensed quasars with small lensing separations are usually 
unresolved in ground-based images and are difficult to distinguish from un-lensed quasars.
One possible way to find these lensed quasars is to identify
quasars with large apparent SMBH masses and the Eddington ratios 
and carry out follow-up high-resolution imaging with {\em HST} or {\em JWST}.
This method has been used in the discovery of the currently only known lensed quasar at $z>5$ \citep{fan19} 
and provided promising lensed quasar candidates (Yue et al. submitted). 

\section{Conclusion} \label{sec:conclusion}

In this paper,
we investigate the strong lensing hypothesis for seven young quasars at $z\gtrsim6$
with lifetimes of $t_Q\lesssim10^5$ yrs, 
identified via their small proximity zone sizes. 
We use high-resolution images taken with {\em HST} 
to search for multiple lensed images of the quasars, 
and use deep images in short wavelengths to detect potential foreground lensing galaxies.
We find no evidence of strong lensing for all seven quasars in our sample,
essentially ruling out the hypothesis that the observed short quasar lifetimes 
are results of strong  lensing.
We further exploit  the distribution of weak lensing magnification 
and derive the impact of lensing magnification on quasar lifetime estimates.
Our main results are:
\begin{enumerate}
    \item The {\em HST} images of these seven quasars are well described by point sources, ruling out lensing models with lensing separations larger than the PSF FWHMs. 
    The strong lensing probabilities of these quasars are estimated to be $\sim1.4\times10^{-5}$.
    
    \item Given the small strong lensing probabilities,
    weak lensing dominates the probability distribution of lensing magnification, $P(\mu)$.
    We compute the probability distribution of $t_Q$ for each quasar 
    by marginalizing all possible values of magnifications.
    {Lensing magnification only shifts 
    the mean values of estimated $t_Q$ by $\lesssim0.2$ dex,}
    and we confirm the short lifetimes ($t_Q\lesssim10^5$ yrs) of the young quasars.
    
    \item The young quasars with $t_Q\lesssim10^5$ yrs are consistent with the picture
    where high-redshift SMBHs have a high obscured fraction,
    have had multiple periods of active accretion,
    and/or have experienced radiatively inefficient super-Eddington accretion phases.
    
    \item We investigate the impact of lensing magnification on measurements 
    of other quasar properties, including the SMBH mass and the Eddington ratio. 
    Such effects should be considered in studies of quasar properties, 
    and provide a viable way to search for compact lensed quasars.
\end{enumerate}


\acknowledgments
{We thank the referee for the valuable comments.}
MY, ACE, and RS acknowledge support by HST-GO-16756 grant from the Space Telescope Science Institute.
CM acknowledges support by the VILLUM FONDEN under grant 37459. 
The Cosmic Dawn Center (DAWN) is funded by the Danish National Research Foundation under grant DNRF140.
JBM acknowledges support by a Clay Fellowship at the Smithsonian Astrophysical Observatory.
These observations are associated with programs GO-13645, GO-15085, and GO-16756.
Some of the data presented in this paper were obtained from the Mikulski Archive for Space Telescopes (MAST) at the Space Telescope Science Institute. 
The specific observations analyzed can be accessed via \dataset[doi:10.17909/y1d0-v310]{http://dx.doi.org/10.17909/y1d0-v310}.

%

\vspace{5mm}
\facilities{{\em HST} (ACS/WFC, WFC3/IR)}


\software{astropy \citep{2013A&A...558A..33A,2018AJ....156..123A},
SciPy \citep{2020SciPy-NMeth}, {\em galfit} \citep{galfit}}




\bibliography{sample631}{}
\bibliographystyle{aasjournal}



\end{document}